\documentclass{aastex}          
\usepackage{spr-astr-addons}    

\newcommand\kms{km~s$^{-1}$}
\newcommand\kmsMpc{km~s$^{-1}\,$Mpc$^{-1}$}
\newcommand\etal{{et al.}} 
\newcommand\mM{\ifmmode(m{-}M)\else$(m{-}M)$\fi}
\newcommand\mMfv{\ensuremath{\Delta(m{-}M)_{FV}}}

\newcommand\hst{{\it HST}}
\newcommand\zacs{\ifmmode z_{850}\else$z_{850}$\fi}
\newcommand\Iacs{\ifmmode I_{814}\else$I_{814}$\fi}
\newcommand\iacs{\ifmmode i_{775}\else$i_{775}$\fi}
\newcommand\gacs{\ifmmode g_{475}\else$g_{475}$\fi}
\newcommand\racs{\ifmmode r_{625}\else$r_{625}$\fi}
\newcommand\vacs{\ifmmode V_{606}\else$V_{606}$\fi}
\newcommand\gz{{\ifmmode{(g_{475}{-}z_{850})}\else$(g_{475}{-}z_{850})$\fi}}
\newcommand\gzacs{\gz}
\newcommand\riacs{{\ifmmode{r_{625}{-}i_{775}}\else$r_{625}{-}i_{775}$\fi}}
\newcommand\rzacs{{\ifmmode{r_{625}{-}z_{850}}\else$r_{625}{-}z_{850}$\fi}}
\newcommand\izacs{{\ifmmode{i_{775}{-}z_{850}}\else$i_{775}{-}z_{850}$\fi}}
\newcommand\vzacs{{\ifmmode{V_{606}{-}z_{850}}\else$V_{606}{-}z_{850}$\fi}}
\newcommand\vi{{\ifmmode{(V{-}I)}\else$(V{-}I)$\fi}}
\newcommand\gI{{\ifmmode{(g{-}I)}\else$(g{-}I)$\fi}}
\newcommand\gIacs{{\ifmmode{(g_{475}{-}I_{814})}\else$(g_{475}{-}I_{814})$\fi}}
\newcommand\Nbar{\ensuremath{\overline{N}}}

\newcommand\Nzbar{\ensuremath{\overline{N}_{z}}}

\newcommand\zbar{\ensuremath{\overline{z}_{850}}}

\newcommand\mtot{\ensuremath{m_{\rm tot}}}
\newcommand\mbar{\ensuremath{\overline{m}}}

\newcommand\Mbar{\ensuremath{\overline{M}}}

\begin{document}
%
\title{Surface Brightness Fluctuations as Primary and Secondary Distance Indicators}

\shorttitle{SBF Distances}
\shortauthors{J.P.~Blakeslee}

\author{John P. Blakeslee\altaffilmark{1}} 
\affil{Dominion Astrophysical Observatory, Herzberg Institute of Astrophysics, National Research Council of Canada, Victoria, BC V9E\,2E7, Canada}

\begin{abstract}
The surface brightness fluctuations (SBF) method measures the variance in a
galaxy's light distribution arising from fluctuations in the numbers and
luminosities of individual stars per resolution element.  Once calibrated for
stellar population effects, SBF measurements with \hst\ provide distances to
early-type galaxies with unrivaled precision.  Optical SBF data from \hst\ for
the Virgo and Fornax clusters give the relative distances of these nearby
fiducial clusters with 2\% precision and constrain their internal structures.
Observations in hand will allow us to tie the Coma cluster, the standard of
comparison for distant cluster studies, into the same precise relative distance
scale.  The SBF method can be calibrated in an absolute sense either empirically
from Cepheids or theoretically from stellar population models.  The agreement
between the model and empirical zero points has improved dramatically,
providing an independent confirmation of the Cepheid distance scale.  SBF is
still brighter in the near-IR, and an ongoing program to calibrate the method
for the F110W and F160W passbands of the Wide Field Camera~3 IR channel 
will enable accurate distance
derivation whenever a large early-type galaxy or bulge is observed in these
passbands at distances reaching well out into the Hubble flow.
\end{abstract}


%


\section{Background}\label{s:intro}

The surface brightness fluctuations (SBF) method remains one of the most accurate ways
of measuring galaxy distances.  \cite{ts88} presented the original implementation of
the method; further developments were described by \cite{tal90}, \cite{jen98},
\cite{bat99}, and \cite{mei05a}.  The method determines the intrinsic variance in a
galaxy image resulting from stochastic variations in the numbers and luminosities
of the stars falling within individual pixels of the image.  These variations in pixel
intensity are blurred by the point spread function (PSF); thus, the variance
measurement is done in Fourier space on the scale of the PSF, and all sources of
contamination (unresolved background or foreground sources) must be taken into account. 
The measured variance is normalized by the local galaxy surface brightness and then
converted to the apparent SBF magnitude~\mbar.  
If the galaxy distance is known, then the absolute SBF magnitude \Mbar\ provides useful
information on the properties of the most luminous stars within the galaxy, and
measurements of \mbar\ in different bandpasses yield ``fluctuation colors'' that provide
distance-independent constraints on the stellar population.  For more about using SBF
for stellar population studies, see recent works by \cite{can07b}, \cite{rai09}, \cite{sbfpops},
\cite{gonz10}, and \cite{lee10}.

For the purpose of estimating galaxy distances, the absolute \Mbar\ must be predicted
accurately from stellar population synthesis models (SBF as a primary distance
indicator) or from an empirical calibration (secondary distance indicator).  The
distance modulus, \mbar$\,{-}$\Mbar\ then follows directly.    As the SBF method has
been around for over two decades, the estimated distances have been compared to those
produced by many other methods, including 
Cepheids \citep{ton97,ton00,lff00},
Type~Ia supernovae \citep{ajhar01},
the fundamental plane \citep{bla02},
the planetary nebula luminosity function \citep{cia02},
and the tip of the red giant branch \citep{mould09}.
However, in most cases, these comparisons involve the ground-based SBF distances from \cite{ton01}.
\cite{bla09,bla10} compare the recent SBF results from the \textit{Hubble Space Telescope}
(\hst) to earlier ground-based work and discuss the reliability and
limits of the older data (see in particular Appendix~A of Blakeslee \etal\ 2010), as
well as zero-point calibration issues.
The purpose of the present contribution is to describe the current state of the SBF art
and foreshadow future developments.
The treatment here is woefully incomplete, and the interested reader should
consult the original works for further details.

\section{SBF Studies with \hst/ACS}

Most of the recent observational work on SBF distances has been carried out with the
Advanced Camera for Surveys (ACS) on board \hst.  Earlier \hst\ studies using WFPC2
were limited by the sensitivity, field size, and/or sampling of that instrument.  The
installation of ACS on \hst\ made the SBF method far more powerful.  The
ACS Wide Field Channel (ACS/WFC) samples the PSF with a resolution comparable to WFPC2's
planetary camera detector (40\arcsec\ field), but over a much larger ${\sim\,}3\farcm3$
field of view and with about five times the throughput at the red wavelengths typically used
for SBF distance measurements.  These characteristics made it possible to measure significant radial
gradients in the SBF amplitude, corresponding to stellar population gradients, for sizable
samples of early-type galaxies  \citep{can05,can07b}.  ACS also vastly improved the
precision of mean SBF magnitudes, and thus of galaxy distances.

\subsection{ACS Virgo Cluster Survey}

The ACS Virgo Cluster Survey (ACSVCS, \citeauthor{cote04} \citeyear{cote04}) is a Large
Program with \hst\ targeting 100 early-type galaxies in the Virgo cluster with the F475W
(\gacs) and F850LP (\zacs) bandpasses.  The goals of the survey include detailed,
high-resolution studies of galaxy nuclei and central structure, the properties of
thousands of globular clusters in the program galaxies, and the structure of the Virgo
cluster as determined by SBF distance measurements.  Here, we concentrate on the 
distance results.

This survey posed a number of challenges for the SBF method.  The off-axis ACS
instrument has significant spatial distortion, and the standard linear interpolation kernel
used for the image rectification had severe implications for the image power spectrum,
making the usual SBF analysis impossible.  As discussed in \cite{mei05a}, the problem
was solved using a damped sinc function for the interpolation kernel.  The calibration
was also a challenge.  The large range in galaxy luminosity (a factor of over 400)
was attended by a large range in color, and with the high precision of the ACS
measurements, the usual linear calibration \citep{ton97,ton01,bva01,jen03,mieske06} 
proved inadequate \citep{mei05b}.  Once these issues were addressed, the quality of the
results were unprecedented, with a mean distance modulus error of 0.07~mag, or about
0.5~Mpc for Virgo.  This enabled the first clear resolution of the line-of-sight depth
of Virgo, as reported by \cite{mei07}.

\begin{figure}[tb]
\includegraphics[width=0.95\columnwidth]{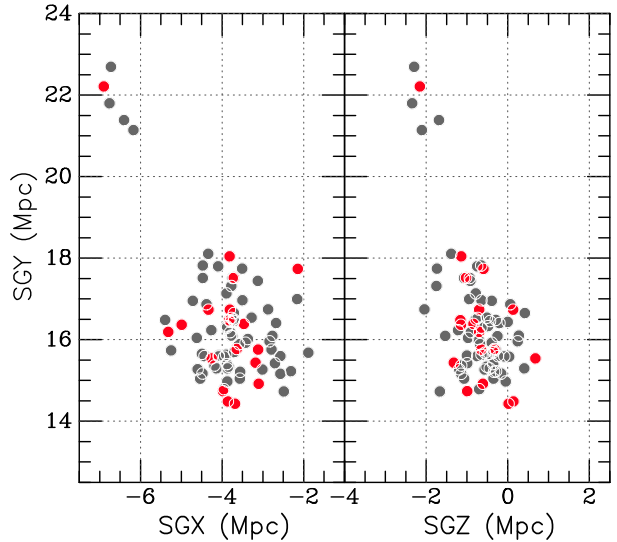}
\caption{ACS Virgo SBF distances in supergalactic coordinates.  Line-of-sight
  distance is roughly along the supergalactic $y$ axis (SGY).  The most luminous
  galaxies (a complete magnitude-limited subsample) are shown in red.
  The 5 galaxies near SGY $\approx22$ Mpc are projected within the Virgo cluster on the
  sky, but are members of a background group centered
  on the giant elliptical NGC\,4365.}
\label{fig:virstruct}
\end{figure}

\begin{figure}[tb]
\includegraphics[width=\columnwidth]{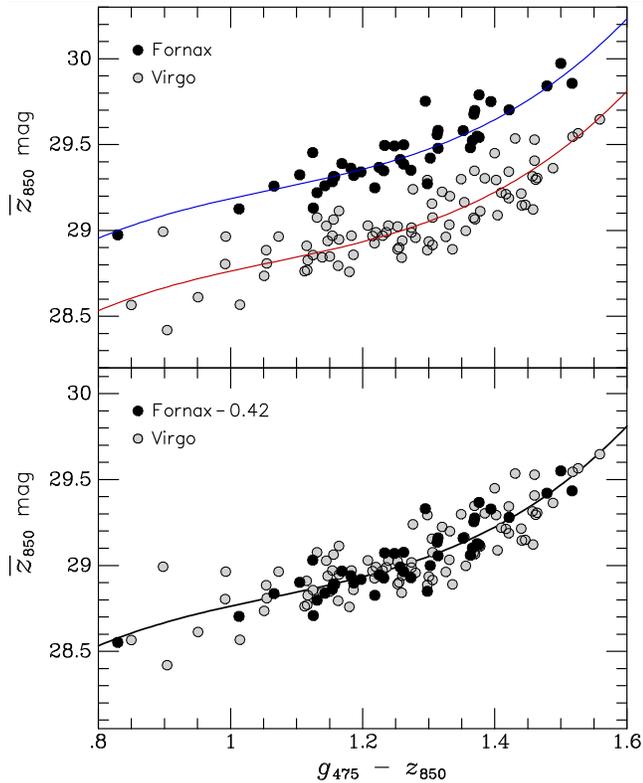}
\caption{SBF \zbar\ magnitude versus \gz\ color for the combined ACS Virgo and
    Fornax samples.   The curves show the best-fit cubic polynomial for the full sample of galaxies,
    assuming the value of the relative Fornax-Virgo distance modulus \mMfv\ that gave the
    minium $\chi^2$. The lower panel shows the two samples shifted together by subtracting
  the best-fit $\mMfv=0.42\pm0.02$ mag from the Fornax galaxy SBF magnitudes.}
\label{fig:finalcal}
\end{figure}

\begin{figure}[tb]
\includegraphics[width=0.98\columnwidth]{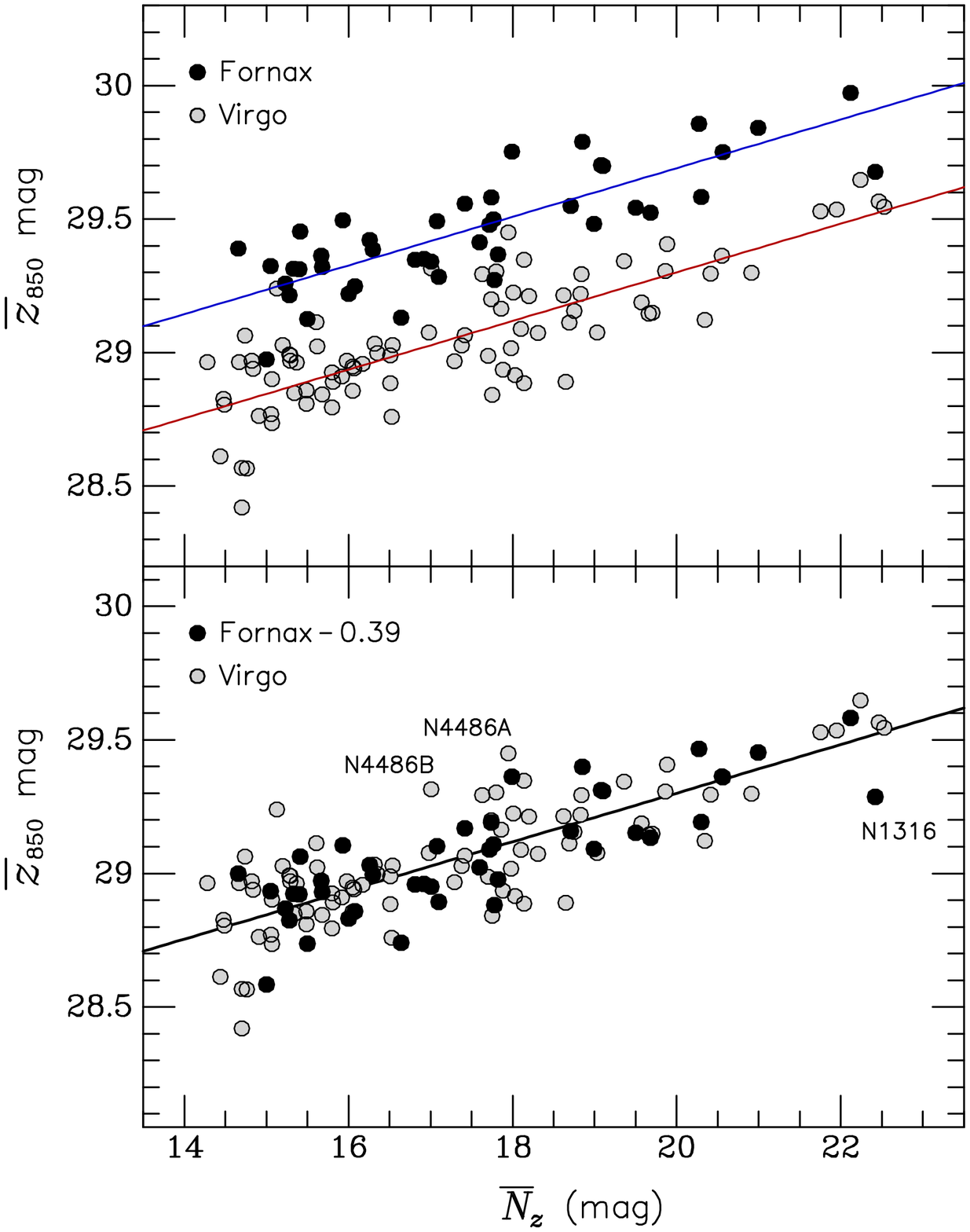}
\caption{SBF calibration similar to Figure~\ref{fig:finalcal}, but based on the
  \zacs-band ``fluctuation count'' \Nzbar.
  A~linear fit works well, but
  the scatter is larger than for the polynomial calibration against \gz.
  The best-fit relative distance modulus $\mMfv=0.39{\pm}0.03$ mag is consistent with that from the
  color calibration, despite the potential for environmental bias in this relation with \Nbar\ (see text).
  The luminous blue post-merger Fornax galaxy NGC\,1316 is a prominent outlier at the
  bright end because its SBF magnitude is ``too bright'' for its luminosity, while
  under-luminous red galaxies such as the tidally stripped companions of
  NGC\,4486 (M87) scatter above the mean relation because their SBF is relatively ``too faint.''
The calibration of SBF against color is immune to such effects.
} 
\label{fig:nbarcal}
\end{figure}

Figure~\ref{fig:virstruct} shows the ACSVCS SBF distances in supergalactic coordinates.
The most obvious feature is the group of 5~galaxies, the brightest being NGC\,4365,
about 6~Mpc beyond the main concentration of Virgo.  These galaxies are within the Virgo
cluster on the sky, and even have the same mean radial velocity as a result of their
infall motion from the far side.  However, the SBF distances clearly reveal them
as a distinct group.  The cluster itself has a triaxial structure with axial ratios
1\,:\,0.7\,:\,0.5, with the smallest extent being in the super\-galactic $z$ direction.  The total
line-of-sight depth is approximately 2.4~Mpc (based on a 0.6~Mpc rms depth).  In
addition to the NGC\,4365 group, another grouping of galaxies associated with NGC\,4406
(M86) is about 1~Mpc beyond the mean cluster distance and has a high (negative) velocity
through the cluster.  The two most luminous galaxies, M87 and M49, both have
distances consistent with the cluster mean; the spatial distribution of dwarf galaxies
is very similar to that of the giants.  For additional results on Virgo structure, see
\cite{mei07}.  Recalibrated distances for the sample galaxies are given in
\cite{bla09}.

\subsection{ACS Fornax Cluster Survey}

The ACS Fornax Cluster Survey (ACSFCS, Jord{\'a}n et al.\ 2007) targeted a
complete sample of 43 early-type galaxies in Fornax, the next nearest cluster after
Virgo.  The goals and observing strategy were very similar to those of the ACSVCS.  The
compact structure of Fornax makes it useful for distance calibrations.  Consequently,
one of the primary goals of the ACSFCS was to refine the ACS \zacs-band SBF calibration,
in addition to deriving the relative distance between the Virgo and Fornax clusters.  An
accurate relative distance is essential for making useful comparisons between these
two fiducial clusters.

\begin{figure}[tb]\epsscale{1.0}
\plotone{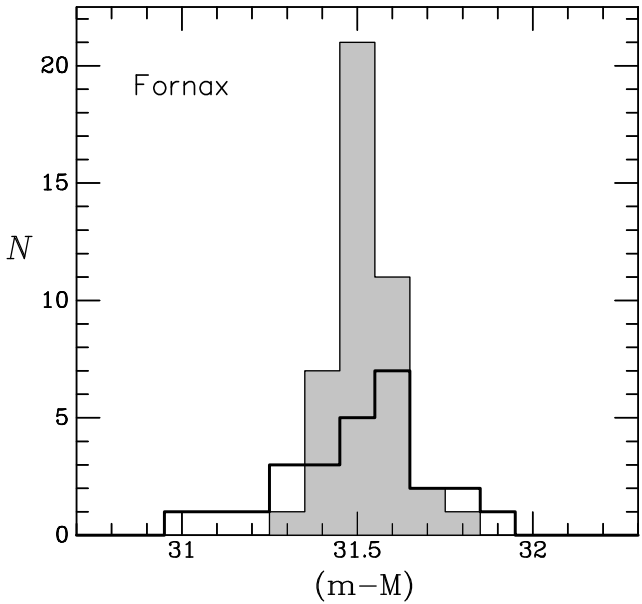}
\caption{Comparison of Fornax galaxy ACS SBF distance moduli from \citeauthor{bla09}
  (\citeyear{bla09}, gray-filled histogram) with Fornax ground-based SBF distances from
  \citeauthor{ton01}  (\citeyear{ton01}, thick-lined open histogram).  SBF measurements
  are possible for many more galaxies using \hst/ACS, and the resulting distance
  distribution is much tighter than in the earlier ground-based data (for which the histogram width
  mainly reflects measurement error).
}\label{fig:fornaxhisto}
\end{figure}

Figure~\ref{fig:finalcal} shows the measured ACS \zacs\ SBF magnitudes as a function of
\gzacs\ colors for both the Virgo and Fornax samples \citep{bla09}.  The curves show the
cubic polynomial calibration fitted to the combined sample (omitting the five ACSVCS
galaxies in the background group).  The offset in magnitudes was varied, and the minimum
$\chi^2$ occurs for a relative Fornax-Virgo distance modulus of $\mMfv = 0.42\pm0.02$
mag.  Thus, the Fornax cluster is 21\% more distant than Virgo, or at 20~Mpc for the
adopted distance zero point.  Although this relative distance is
consistent with some previous studies, the precision here is far higher.  The intrinsic
scatter in the SBF versus color relation is only 0.06~mag for the galaxies in this
sample with $\gz>1$.  The calibration ``flares out'' at bluer colors, although this is
mainly evident for the Virgo sample and could be partly the result of distance
variations.   Unlike the situation for Virgo, 
there is no evidence for significant substructure in the Fornax
cluster from the ACSFCS sample.

\cite{bla09} also presented an alternative SBF calibration based on the `fluctuation
count' parameter $\Nbar = \mbar - \mtot$, first introduced by
\cite{ton01} and tested for calibration of the ground-based data
by \cite{bla02}, where $\mtot$
is the total apparent galaxy magnitude.  \Nbar~corresponds to the luminosity of the
galaxy in units of the luminosity-weighted mean stellar luminosity.  It is therefore a
distance-independent measure of the galaxy mass and correlates with \mbar\ as a
result of the galaxy mass--color relation.  
Figure~\ref{fig:nbarcal} shows the calibration based on \Nbar, which is approximately
linear and yields a consistent relative distance (within the uncertainties) but with a
larger intrinsic scatter of $\sim0.10$~mag.

Unlike the calibration based on color, which is based purely on stellar population
properties, the relation with \Nbar\ involves scaling
relations similar to the fundamental plane.  Thus, luminous blue galaxies, which may have experienced recent
star formation and therefore have SBF magnitudes ``too bright'' for their large 
luminosities, and small red galaxies, which may be tidally stripped and therefore have
SBF magnitudes ``too faint'' for their luminosities, will deviate from
the mean relation.  Moreover, the \Nbar\ relation is likely to have an environmental dependence. 
For example, if the galaxies in Fornax are younger on average than those in Virgo, the
distance offset from the \Nbar\ calibration would be underestimated. For these reasons, the
calibration based on color is preferable, but \Nbar\ provides an
alternative when an appropriate color is unavailable.  For example, \cite{can11} have used a
calibration based on \Nbar\ to derive distances for 12 galaxies observed
with the VLT.  Moreover, \Nbar\ is an interesting quantity in its own right, and can be
used to construct distance-independent color-magnitude diagrams.

\subsection{An Improved $I$-band Calibration and Comparisons to Ground-based SBF Distances}

Figure~\ref{fig:fornaxhisto} shows a comparison of the color-calibrated ACSFCS SBF
distances from \cite{bla09} to the ground-based Fornax SBF distances from \cite{ton01}.
The two surveys agree well in the mean, but the distribution of distance moduli has an
rms dispersion of 0.092~mag for the ACS data, compared to 0.21~mag for the ground-based
SBF data.  These dispersions include contributions from intrinsic scatter
($\sim0.06$ mag) and the depth of Fornax ($\sim0.05$ mag, estimated from the
distribution on the sky), which are the same for the ACS and ground-based data.  Thus,
the actual measurement errors are about 4~times smaller for the ACS data.  The reasons
for this vast improvement include a much stronger signal due to the sharper PSF, the
ability to identify and remove contaminants to much fainter levels, the improved
stability of the PSF, and the photometric homogeneity of the ACS data. The ACS/WFC
on \hst\ is an almost ideal instrument for optical SBF measurements, although a still larger
field would allow better sky estimation for large galaxies.

\begin{figure}[tb]\epsscale{1.0}
\plotone{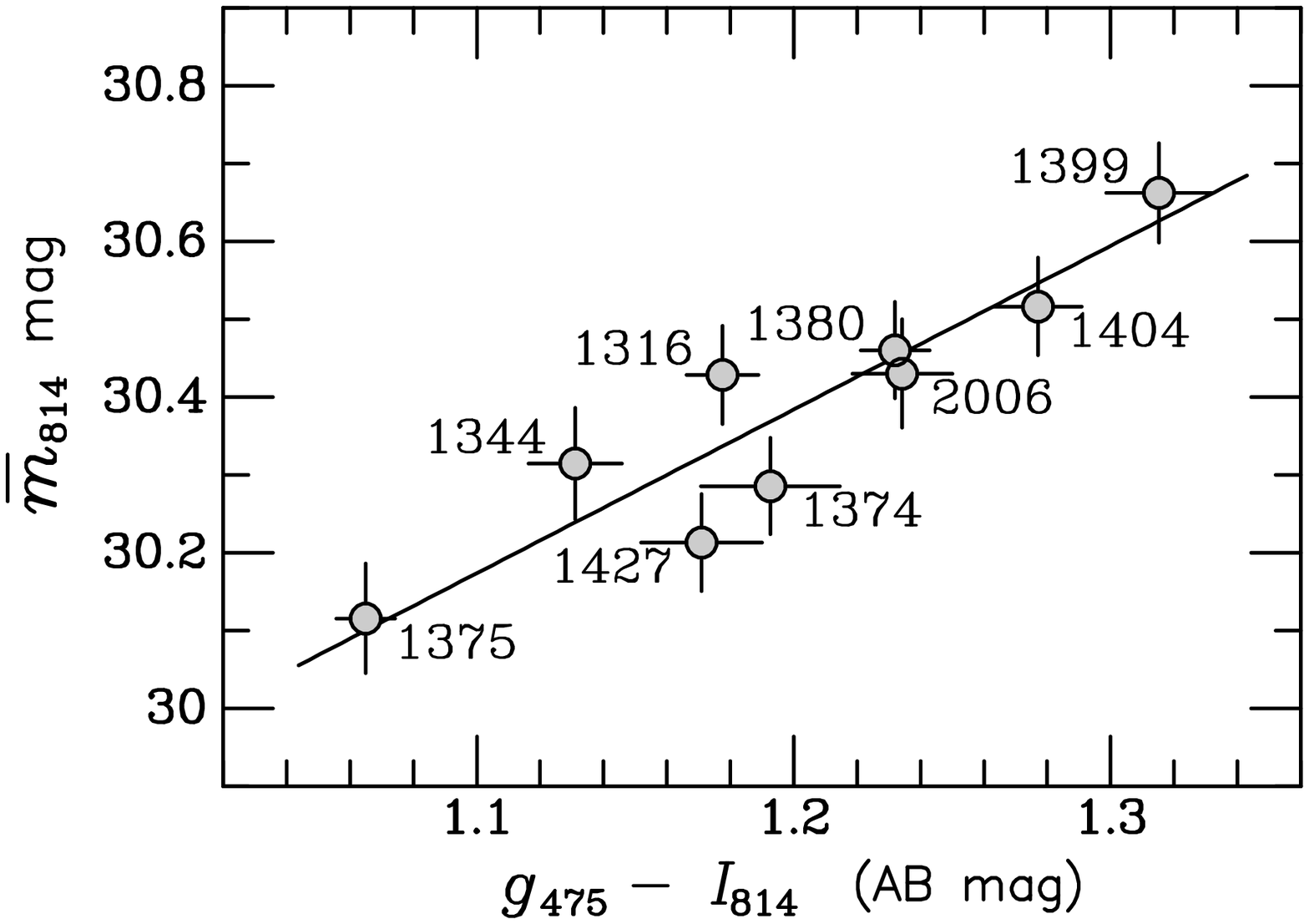}
\plotone{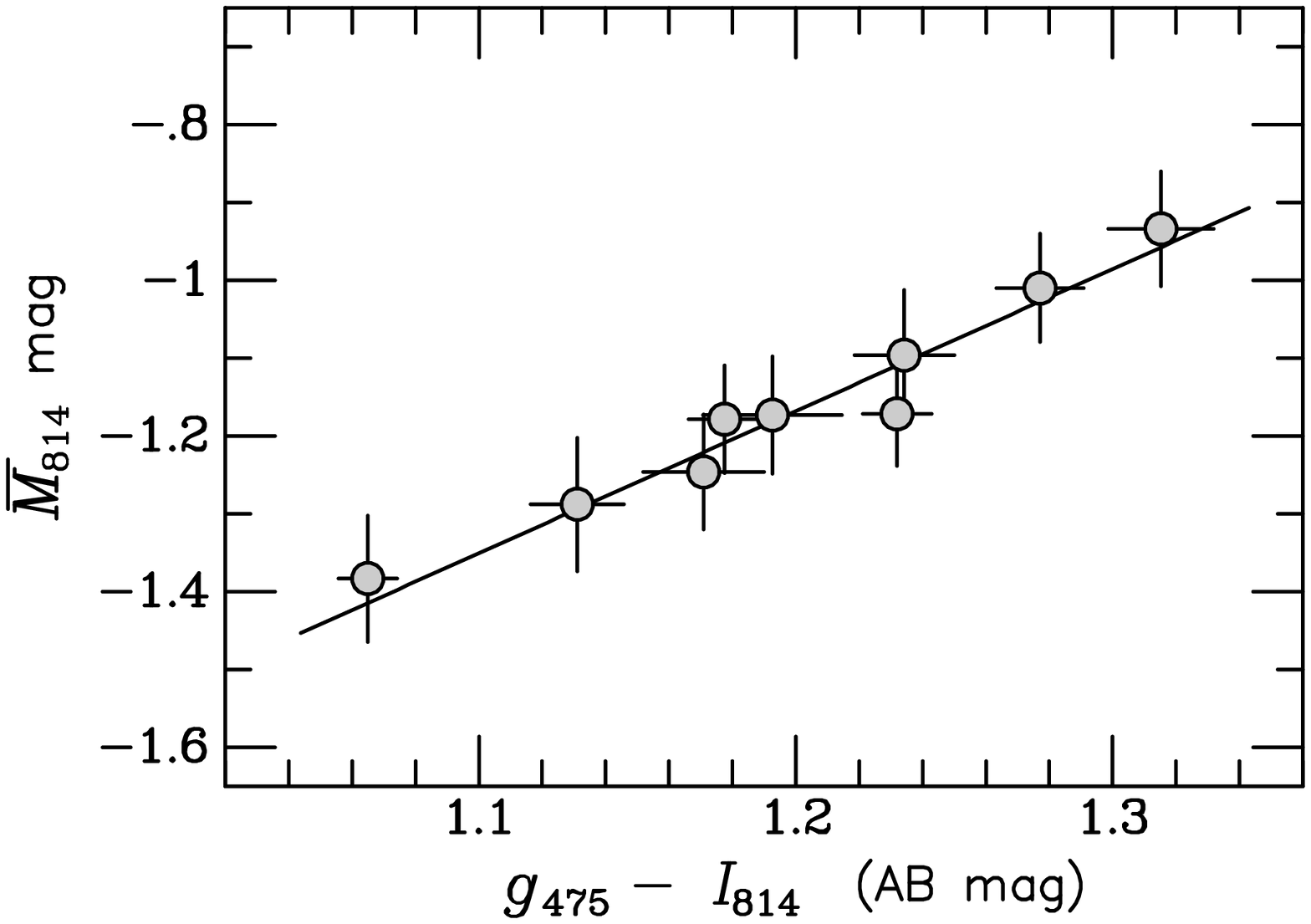}
\caption{
The F814W SBF calibration.  The top panel shows the measured apparent 
SBF magnitudes as a function of \gIacs\ color for nine Fornax galaxies, labeled by their
NGC or IC numbers.
The lower panel shows absolute SBF magnitudes using the \zacs\ SBF distance moduli from
the ACSFCS.  The lines indicate linear fits, with the coefficients reported by \cite{bla10}.
The very small scatter in the lower panel results from the intrinsic correlation between
SBF measurements in the very similar \Iacs\ and \zacs\ bandpasses.
}\label{fig:isbfcal}
\end{figure}

The ACS Virgo and Fornax surveys used the \zacs\ bandpass for SBF measurements,
as noted above.  Although this provided excellent distances with only one orbit per
galaxy, \zacs\ is not the most efficient filter. 
The F814W bandpass (\Iacs) is similar to the Cousins $I$-band used
for the ground-based SBF survey \citep{ton97,ton01} and has a throughput more than twice
that of \zacs, making \Iacs\ preferable for SBF studies of more distant galaxies that would
require multi-orbit observations.  Although there had been some \Iacs\ SBF studies that
used a transformed version of the ground-based calibration 
\citep{can07a,can07b,gina07} or a theoretical calibration \citep{biscardi08}, an
empirical calibration for the ACS \Iacs\ band was needed for improved reliability.

Figure~\ref{fig:isbfcal} presents the empirical \Iacs\ SBF calibration based on 9~bright
early-type galaxies in the Fornax cluster \citep{bla10}.  The top panel shows the
relation between apparent SBF magnitude and \gIacs\ color; the scatter is
0.06~mag.  When the measured \zacs\ SBF distances from the ACSFCS are used to derive
absolute \Iacs\ SBF magnitudes, the resulting relation shown in the lower panel has a
scatter of only 0.03~mag.  This small scatter, which is less than expected from
stellar population variations at a given color, results because the intrinsic scatter
in the SBF magnitudes is correlated for the very similar \Iacs\ and \zacs\ bands.  
This helps to ensure consistency in the distances derived from data in these two bands;
see the original work for further discussion.  
A linear fit to the calibration in Figure~\ref{fig:isbfcal} suffices because the
galaxies are all relatively red, and the curvature is only expected to become
significant at bluer colors.

A major part of this ACS \Iacs\ SBF calibration study included further comparisons between the
ground-based SBF distances from \cite{ton01} and the \hst/ACS SBF distances.  The
results are given in Appendix~A of Blakeslee et al.\ (2010), which includes a correction formula, based
on the \cite{ton01} ``quality'' index $Q$, for removing a small apparent bias in the
ground-based SBF distance compilation.   In addition, an offset is derived to bring the 
\cite{jen03} NICMOS SBF distances into consistency with the ACS and corrected ground-based
distances. The interested reader should refer to that Appendix for a detailed discussion
of this important issue.

\begin{figure}[tb]
\plotone{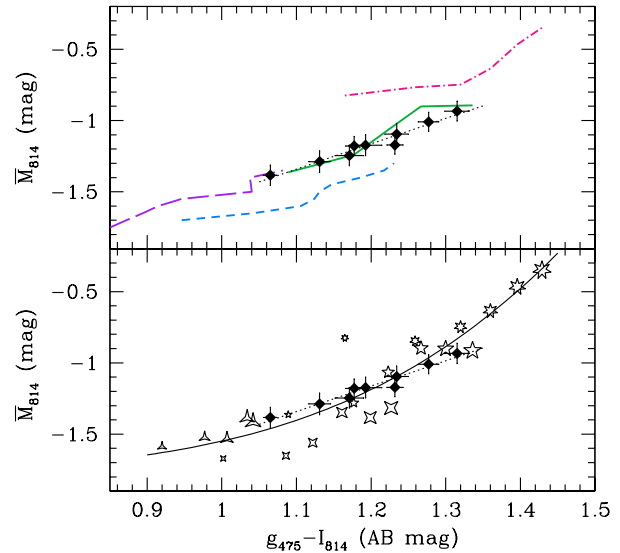}
\caption{Calibrating SBF as a primary distance indicator.
The top panel shows predicted curves for Teramo/SPoT models with metallicities
$[\hbox{Fe/H}] = -0.7, -0.3, 0.0, +0.3$ (long-dashed purple, short-dashed blue, solid
green, and dot-dashed red curves, respectively) 
and ages from 3 to 14~Gyr.  The solar metallicity model predictions follow 
the data points (black diamonds) and empirical linear relation (dotted line),
without adjustment in zero point.
The bottom panel is similar to the top, but here the individual models are represented by
discrete points, with different symbols for different metallicities, and symbol size
increasing with age.
The curve is a cubic polynomial fit to the plotted models; within the color range of the
data points, the relation is fairly linear, but the models predict increased
curvature beyond this range, similar to what has been observed for the \zacs-band
over a larger empirical color range.}
\label{fig:models}
\end{figure}

\section{SBF as a Primary Distance Indicator}

Because the absolute SBF magnitude in a given bandpass is an intrinsic property of a
stellar population, it is possible to calibrate the SBF method from theoretical
modeling, and there are many examples in the literature
\citep{tal90,wor93,liu00,bva01,can03,marin06,biscardi08,gonz10}. 
Calibrations based on simple stellar population models are complicated when the effects
of age and metallicity are not fully degenerate, since it is often unclear which mix of
models best match real galaxies.  In particular, this makes any theoretical estimation of the
intrinsic scatter very difficult.  However, model comparisons are useful as a sanity
check, at the very least.  \cite{mei05b} showed that their empirical \hst/ACS \zacs\ SBF
calibration agreed reasonably well with the predictions from the \cite{bc03} models,
although with a zero-point offset of $\sim0.15$~mag.  It would be interesting to update
this comparison with the newer version of these models that is now available.

Figure~\ref{fig:models}, from \cite{bla10}, compares the \hst/ACS \Iacs\ SBF data and
empirical calibration to the predictions from the Teramo SPoT models \citep{rai05,rai09},
as reported by \cite{biscardi08}, but transformed to the AB system.  The linear
empirical calibration matches remarkably well with the predictions from the SPoT solar
metallicity models, without any adjustment in the empirical or model zero points.
The fitted relations to data and models intersect near the mean color of
$\gIacs = 1.2$.  As the empirical SBF zero point here is tied to Cepheids \citep{mei07,bla10},
the consistency with the stellar population model predictions can be taken as an
affirmation of the empirical distance scale, which locates the Virgo and Fornax clusters
at 16.5~and 20.0~Mpc, respectively.  Thus, using SBF as a primary distance indicator,
with model calibration given by the polynomial fit in the lower panel of
Figure~\ref{fig:models}, agrees very well with the standard SBF secondary distance indicator.

An example of the use of SBF as a primary distance indicator is given by Biscardi \etal\
(2008), who measured SBF distances for 4~galaxies with radial velocities in the 4000 to
8000~\kms\ range, and thus distances extending beyond 100~Mpc.  These authors used a
theoretical calibration based on the Teramo SPoT models, and obtained a value for the
Hubble constant $H_0 = 76\pm6$ \kmsMpc, where the error bar here reflects the
statistical uncertainty.  The systematic uncertainty due to the model zero point is
harder to quantify, but the agreement with the best current values for $H_0$ 
based on the empirical distance ladder \citep{fm10,rie11}  is highly encouraging.

\begin{figure*}[th]
\begin{center}
\includegraphics[scale=0.222]{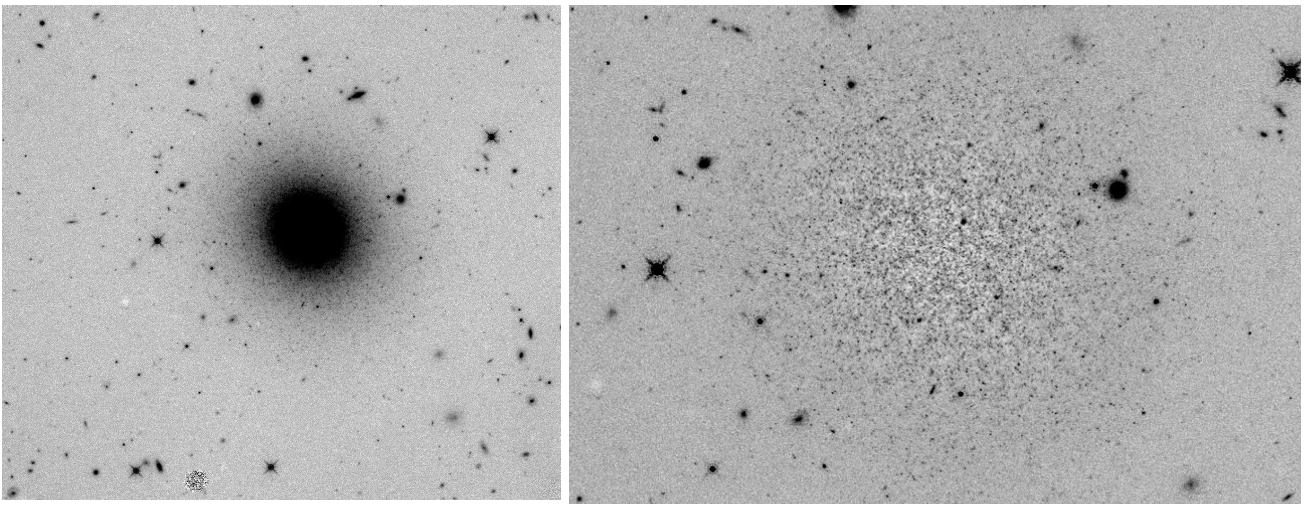}
\end{center}
\caption{WFC3/IR  image of the Virgo galaxy IC\,3032 in the F160W bandpass (left); an enlarged
  view of the image after  galaxy model subtraction (right). The fluctuations are evident.} 
\label{fig:ic3032}
\end{figure*}

\section{SBF with WFC3/IR}

Although the recent work on SBF in the optical from the ACS and VLT is extremely
impressive when compared to earlier ground-based and pre-ACS \hst\ studies, perhaps the
most exciting prospects for future SBF work lies at near-IR wavelengths.  Such
statements as this have been commonplace for well over decade (e.g., Blakeslee \etal\
1999), but now, with the installation of the Wide Field Camera~3, with its IR channel
(WFC3/IR), on \hst, the payoff finally appears imminent.  The SBF signal is typically an
order of magnitude brighter in the near-IR than in the optical, and the characteristics
of WFC3/IR make it far more efficient for SBF work than previously available near-IR
cameras.  A current \hst\ program is carrying out a calibration of the SBF method for
the F110W and F160W bandpasses of WFC3/IR using observations of 16 Virgo and Fornax
cluster galaxies that already have good ACS SBF distance measurements.  The new data
appear spectacular, with some results on globular clusters from this program presented
by \cite{bla11}.  The WFC3/IR F160W image of the Virgo galaxy IC\,3032 is shown in
Figure~\ref{fig:ic3032}, before and after galaxy model subtraction.  The fluctuation
signal is extremely high.

The combination of the bright near-IR SBF signal with the high throughput and wide field 
of WFC3/IR should make it possible to obtain an accurate SBF distance whenever an
early-type, or bulge-dominated, galaxy is observed in the calibrated passbands at
distances reaching beyond 150~Mpc, well out into the Hubble flow.  Eventually, once a
large number of galaxies have been observed across the sky with WFC3/IR, SBF
measurements from archival data could enable highly detailed mapping of the mass density
distribution in the local universe.  However, some aspects of the WFC3/IR SBF analysis
require further testing, and the intrinsic scatter in the SBF calibration at these near-IR
wavelengths still needs to be quantified with the same precision as for the ACS F814W
and F850LP bandpasses.  The results from this effort are expected in the near future!

\section{Summary}\label{s:sum}

The SBF method has been in use for over two decades.  The survey of 300 galaxy
distances published by \cite{ton01} has proven to be a valuable resource for numerous different
studies requiring large numbers of nearby galaxy distances.  However, the reliability
and precision of that survey was limited by the heterogeneous, and sometimes barely
adequate, quality of those 1990s-era ground-based data.  During the same period, some
SBF efforts were made with WFPC2 on \hst, but that instrument was far
from ideal for SBF measurements.  However, the installation of the ACS on \hst\ ushered in a
renaissance for optical SBF studies.  To date, \hst/ACS SBF distances have been published for 136
galaxies in the directions of the Virgo and Fornax clusters, compared to fewer than 60
in the ground-based SBF survey.  The distance errors (including intrinsic stellar
population scatter in the method, but omitting systematic zero-point uncertainty)
for the ACS SBF measurements are typically 4-5\%, as compared to 10\% or more for the
ground-based SBF measurements.  The measurement errors (omitting intrinsic scatter) are
several times smaller for the ACS samples, and this has allowed quantification of the
quality-related bias in the ground-based SBF distances.

We now have an excellent calibration for the SBF method in the F814W and F850LP
bandpasses of ACS, and this has allowed us to probe the internal structure of the Virgo
cluster and measure the Fornax/Virgo distance ratio with unprecedented precision.
Recent stellar population models support the empirically based distance zero point and
allow SBF to be applied with confidence as a primary distance indicator.  
With the proper calibration and control of systematics, the space-based SBF method 
is the most accurate way of measuring the distances to early-type galaxies in the
important $\sim10$ to $\sim100$ Mpc distance range.  It is straightforward at smaller
distances, but resolved stellar population methods are then competitive, and at larger
distances the peculiar velocities are generally a small fraction of the Hubble
velocity.   This leaves a large volume for SBF studies.  Additional
data are in hand and efforts are ongoing to bring the Coma cluster into the tight ACS
SBF distance ladder and to probe structures at even larger distances.

The installation of the WFC3 instrument with its IR channel on HST promises to engender
a similar renaissance for the near-IR SBF method as ACS did for optical~SBF. In fact,
the relative advance will likely be even greater, given the vastly lower sky background for
near-IR observations from space.  We have shown the high quality of the WFC3/IR data.
Our efforts now are focused on producing an SBF calibration for WFC3/IR
comparable to that obtained for ACS and characterizing the internal scatter in the
method at these wavelengths.  Given the very modest exposure times required for near-IR
SBF measurements, the availability of such a calibration will greatly improve the
potential for distance and velocity studies from the large cache of \hst\ imaging data.
Much work remains to be done in this area: as the saying goes, 
``The harvest is plentiful, but the workers are few.''  The best is yet to come.

\acknowledgments
I am indebted to my sundry SBF collaborators, including my \hbox{colleagues}
from the ACS Virgo and Fornax cluster \hbox{surveys,} for their many
contributions to the work described here.  
This paper is based on an invited review talk at the conference, ``Fundamental
Cosmic Distance Scale: State of the Art and the Gaia Perspective.''  I thank
the conference organizers for an excellent meeting and for inviting me to talk
on this subject.

%

\end{document}